\documentclass[prl,showpacs,twocolumn,floatfix]{revtex4}

\usepackage{amsmath,amssymb}
\usepackage{graphicx}

\usepackage{color}

\renewcommand {\d} {{\rm d}}
\renewcommand {\i} {{\rm i}}

\renewcommand {\Re} {{\rm Re}}

\newcommand {\ee} {{\rm e}}

\newcommand {\E}  {{\varepsilon}}
\newcommand {\om}  {{\omega}}
\newcommand {\Om}  {{\Omega}}

\newcommand {\bfa} {{\bf a}}
\newcommand {\bfd} {{\bf d}}
\newcommand {\bfe} {{\bf e}}
\newcommand {\bfp} {{\bf p}}
\newcommand {\bfr} {{\bf r}}

\newcommand {\rmA} {{\rm A}}
\newcommand {\rmC} {{\rm C}}

\newcommand {\amax} {a_{\rm max}}
\newcommand {\ex} {{\rm ex}}

\newcommand {\calM} {{\cal M}}

\newcommand {\hatD} {{\widehat{d}}}

\begin{document}

\title{Confinement resonances in photoionization 
of endohedral atoms: a myth or reality?}

\author{A.~V.~Korol}
\author{A.~V.~Solov'yov}
\affiliation{
Frankfurt Institute for Advanced Studies,
Goethe Universit\"{a}t,\\
Ruth-Moufang-Str.~1, 60438 Frankfurt am Main, Germany}

\begin{abstract}
We demonstrate that the structure of confinement resonances 
in the photoionization cross section of an endohedral atom 
is very sensitive to the mean displacement
$\langle a \rangle $ of the atom from the cage center.
The resonances are strongly suppressed if $2\langle a \rangle$
exceeds the photoelectron half-wavelength. 
We explain the results of recent experiments which contradict the 
earlier theoretical predictions on the existence of 
confinement resonances in particular endohedral systems.

\end{abstract}

\pacs{32.80.Fb, 32.90.+a, 36.40.-c}

\maketitle
 
In this Letter we explain the discrepancy between theoretical 
predictions and experimental
results for the photoionization (PI) of endohedral atoms A@C$_{N}$.
The discrepancy concerns 
'confinement resonances' \cite{ConneradeDolmatovManson1999} 
in the PI spectrum whose existence 
was predicted theoretically 
\cite{PuskaNieminen1993,WendinWaestberg1993,LuberekWendin1996,Baltenkov1999,DeclevaEtAl1999,AmusiaBaltenkovBecker2000,BaltenkovDolmatovManson2002,Aichinger2003,BaltenkovDolmatovEtAl2003,Angular,AmusiaBaltenkovEtAl2005,DolmatovManson2006,BaltenkovMansonMsezane2007,DolmatovBrewerManson2008,ChenMsezane2009,Dolmatov2009}
(and references therein) but so far has been not supported experimentally 
\cite{Mitsuke_EtAl2005,KatayanagiEaAl_2008,MuellerEtAl2007}.
We also formulate the criterion which allows
one to estimate, for a particular endohedral system, 
the interval of photon energies where the resonances can appear.  

Confinement resonances appear 
as a result of interference between a direct wave of the photoelectron
escaping the atom and the waves due to scattering from the atoms
of the cage.
Depending on the photoelectron momentum
the interference can be constructive 
or destructive.
Thus, the spectrum of the encaged atom acquires additional 
oscillations as compared to the free atom.
It was noted  \cite{PuskaNieminen1993} that the cage-induced
oscillations have the same nature as the extended X-ray absorption 
fine-structure (EXAFS) for solid-state systems.

To date, a number of theoretical investigations have been carried out 
of the features of confinement resonances for various endohedral
systems.
The resonances were studied for noble-gas 
\cite{PuskaNieminen1993,Baltenkov1999,AmusiaBaltenkovBecker2000,AmusiaBaltenkovEtAl2005,DolmatovBrewerManson2008}
and metal 
\cite{PuskaNieminen1993,WendinWaestberg1993,LuberekWendin1996,Baltenkov1999,DeclevaEtAl1999,BaltenkovDolmatovManson2002,Aichinger2003,ChenMsezane2009}
atoms encaged in C$_{60}$, 
for Ar encaged in larger fullerenes  C$_{240}$ and  C$_{540}$
as well as in onion-like structures C$_{60}$@C$_{240}$ etc 
\cite{DolmatovBrewerManson2008},
for hollow atoms \cite{BaltenkovMansonMsezane2007}.
A strong distortion of the atomic 4d-f giant dipole resonance 
due to the scattering from the cage was predicted
for Ba@C$_{60}$ \cite{PuskaNieminen1993,WendinWaestberg1993},
La@C$_{60}$ \cite{PuskaNieminen1993},
Xe@C$_{60}$ \cite{AmusiaBaltenkovEtAl2005} and Ce$^{3+}$@C$_{60}$
\cite{ChenMsezane2009}.
The confinement resonances were also investigated for the angular 
distribution of photoelectrons
\cite{DeclevaEtAl1999,BaltenkovDolmatovEtAl2003,Angular}.
Specific features of the resonances in the PI of endohedral ions 
A@C$_{60}^{\pm z}$ were studied in Ref. \onlinecite{DolmatovManson2006}. 
These and other closely related topics can be found in the recent review 
\cite{Dolmatov2009}.

The experimental data on the PI of A@C$_N$ are sparse due to 
'\dots the difficulty to produce sufficient amounts of purified 
endohedrals for the
gas phase experiments' \cite{MuellerEtAl2007}.  
So far the PI cross section in the region of the 4d-f giant
resonance were measured for 
Ce@C$_{82}$ \cite{Mitsuke_EtAl2005},
Pr@C$_{82}$ \cite{KatayanagiEaAl_2008}
and 
Ce@C$_{82}^{+}$ \cite{MuellerEtAl2007}.
In \cite{MuellerEtAl2007} it is explicitly  stated that
no confinement resonances were detected.
With respect to \cite{Mitsuke_EtAl2005,KatayanagiEaAl_2008} 
one draws this conclusion matching the measured data
to big absolute uncertainties.

Most of the theoretical work has been devoted to the 
confinement resonances in spherical endohedral systems, 
in which the atom is placed at the center of a spherical cage.
The correlation between the confinement resonances
and the off-the-center position of the atom was reported 
for Ba@C$_{60}$ \cite{LuberekWendin1996}, 
Mg@C$_{60}$ and Mg@C$_{60}^{\pm}$ \cite{Aichinger2003}. 
A strong sensitivity of the photoelectron angular distribution 
and total cross section of PI on the displacement 
of the atom from cage center  was demonstrated 
\cite{BaltenkovDolmatovManson2002,BaltenkovDolmatovEtAl2003}.
Specific results were presented for Li@C$_{60}$ and Ar@C$_{60}$.

In  the cited papers the influence of the off-the-center position 
was analyzed for the fixed-in-the-space endohedral fullerenes,
i.e. fixing both the direction and magnitude of the displacement vector
$\bfa$ between the atom and the center.
In experimental conditions either one of these parameters or both 
(depending on the endohedral system) cannot be controlled.
As a result, proper averaging procedures must be adopted in order to 
bring theoretical predictions closer to the observable quantities. 

For example, in metallofullerenes  M@C$_N$ (M stands for a metal atom), 
the atom, as a rule, is located off the center 
(see, e.g., Ref.~\onlinecite{Shinohara2000}).
The equilibrium can be very stable due to a strong bond of M to the 
carbon atoms.
Hence, one can fix the distance $a=|\bfa|$. 
To explore the properties of an individual metallofullerene 
the direction of $\bfa$ also can be  fixed.
However, to compare with the gas phase experiments, any observable, 
including the cross section, must be averaged over the directions.
In this connection we mention Ref. \cite{DeclevaEtAl1999} 
where the atomic 1s PI of M@C$_{60}$ molecules (M=Li, Na, K) 
was theoretically investigated for the gas of fullerenes.
The authors, indicating the off-the-center position of the atom,
ignore its influence on the cross section and carry out the
study placing the atom at the center for 'computational 
simplicity' (see the end of Introduction).
In what follows we demonstrate, that, generally, such an assumption 
is not correct and can lead to erroneous predictions on the 
existence of the confinement resonances.

Other complexes of current interest are endohedral noble-gas 
atoms \cite{Dolmatov2009}.
In these systems the atom stays neutral, and its dynamics is 
defined by the Lennard-Jones interaction with the cage atoms 
\cite{Albert_EtAl2007,Pyykko_EtAl2007}.
The equilibrium position and the amplitude of thermal 
vibrations depend on the atom type, the cage size  and the temperature.
For such complexes it is reasonable to carry out
also the averaging over the $a$ values.

As  we demonstrate below, in many cases the averaging 
destroys the confinement resonance structure. 
Prior to introducing the formalism, let us present
qualitative arguments.
For simplicity, we ignore multiple scattering from the cage.
Then, the oscillations in the cross section are due to the 
interference of the two waves of photoelectron, the direct and 
the scattered ones, both originating from the same source, 
-- the encaged atom.
The atom at the center can be treated as a point-like source.
For $\bfa \neq 0$ the source acquires a 
size $D \approx 2\langle a\rangle$ 
where $\langle a\rangle$ is the mean distance from the center. 
The finite size of the source influences the interference pattern.
When $D$ exceeds the half-wavelength the pattern is destroyed.
Hence, one can expect that for the photoelectron momenta 
\cite{AtomicUnits} 
\begin{equation}
p > p_{\min}=\pi/D
\label{condition}
\end{equation}
the confinement resonances disappear.
The corresponding range of photon energies is 
$\om > \om_{\min} =  p_{\min}^2/2 + I_0$, 
where $I_0$ is the ionization potential of the atomic shell.

To start with the formalism let us note that, since the 
confinement resonances are due to the interference phenomenon,
one can adopt the simplest model for electron-fullerene interaction which
ensures the interference.
We model this interaction with a  $\delta$-like potential well 
\cite{AmusiaBaltenkov1998,Baltenkov1999}.
Choosing the origin at the atom and introducing the position vector 
$\bfa$ of the cage center, one writes:
\begin{equation}
U_s(\bfr;\bfa) = -A\, \delta\left(|\bfr - \bfa| - R\right),
\label{CagePotential.1}
\end{equation}
where $A$ can be related 
to the cage radius $R$ and the electron affinity of the fullerene 
\cite{AmusiaBaltenkov1998,Baltenkov1999}.
For the central position of the atom 
eq. (\ref{CagePotential.1}) defines the 
spherically-symmetric potential $U_s(r)=-A\delta(r-R)$
 used in the cited papers.

The wavefunction $\Psi^{(-)}_{\bfp}(\bfr; \bfa)$  of the photoelectron 
emitted from the encaged atom satisfies the equation \cite{Messiah}:
\begin{eqnarray}
 \!\!
\Psi^{ (-)}_{\bfp}(\bfr; \bfa)
\!
= 
\!
\psi^{(-)}_{\bfp}(\bfr)
-
\!\! 
\int 
\!\!
 \d \bfr^{\prime}
G_{\E}(\bfr, \bfr^{\prime})
U_s(\bfr^{\prime}; \bfa)
\Psi^{(-)}_{\bfp}(\bfr^{\prime}; \bfa).
\label{wf.3}
\end{eqnarray}
The superscript $(-)$ indicates that the asymptotic form 
is 'plane wave + incoming wave', $\bfp$ and $\E=p^2/2$
are the momentum and energy, 
$\psi^{(-)}_{\bfp}(\bfr)$ and $G_{\E}(\bfr, \bfr^{\prime})$ are
the wavefunction and  the retarded Green function
of the electron escaping from the free atom.
 
For an {\em at-the-center}  atom eq. (\ref{wf.3}) is solved by expanding
$\Psi^{(-)}_{\bfp}(\bfr;0)$,
$\psi^{(-)}_{\bfp}(\bfr)$ and $G_{\E}(\bfr, \bfr^{\prime})$
in series over spherical harmonics.
Then, evaluating the integral and matching both sides 
of the equation at $r=R$, one derives
\begin{eqnarray}
\Psi^{(-)}_{\bfp}(\bfr; 0)
=&&
{2\pi \over pr} 
\sum_{lm}
\i^l
\ee^{-\i\delta_l}
\Bigl(
P_{pl}(r)
-
{\alpha P_{pl}(R) g_{pl}(r,R)\over 1 + \alpha g_{pl}(R,R)}\,
\Bigr)
\nonumber\\
&&\times
Y_{lm}(\Om_\bfr)Y_{lm}^{*}(\Om_\bfp),
\label{WF_central.6}
\end{eqnarray}
where $\alpha= 2\pi A$, and $Y_{lm}$ are spherical harmonics.
The radial wavefunctions $P_{pl}(r)$ and the 
scattering phaseshifts $\delta_l\equiv \delta_l(p)$
correspond to the motion in the field of atomic residue. 
The radial Green function $g_{pl}(r,r^{\prime})$ 
is proportional to the product 
of the regular $P_{pl}(r)$ and irregular $\chi_{pl}(r)$
(at $r=0$) solutions of the radial  Schr\"{o}dinger equation:
$g_{pl}(r,r^{\prime}) = - P_{pl}(r_{<})\, \chi_{pl}(r_{>})/2p$
with $r_{>}/r_{<}$ being the largest/smallest of $r, r^{\prime}$
(see, e.g., \cite{Messiah}). 

Using (\ref{WF_central.6}) one relates the amplitude and cross section 
of PI of the endohedral atom to those of the free atom. 
For the free atom, the partial amplitude $M_{ll_0}^{\rmA}$ of 
the dipole transition from the subshell $\nu_0=(n_0l_0)$ 
($n_0$ is the principal quantum number, $l_0$ is the orbital momentum)
is proportional to the matrix element of the radial dipole operator $\hatD$:
$M_{ll_0}^{\rmA}\propto 
\int_0^{\infty} P_{pl}(r)\,\hatD\, P_{\nu_0}(r)$
where $P_{\nu_0}(r)$ is the shell wavefunction and
$l=l_0\pm 1$ due to the dipole selection rules.
To evaluate the  amplitude for the encaged atom one substitutes 
$P_{pl}(r)$ with the expression in the brackets in (\ref{WF_central.6}).
Additionally, assuming the shell radius to be much smaller than $R$, 
one sets $P_{pl}(R)\,g_{pl}(r,R) =P_{pl}(r)\,g_{pl}(R,R)$
and derives 
$M_{ll_0}^{\rmA @\rmC_N}(\bfa=0)\approx 
M_{ll_0}^{\rmA}/(1 + \alpha g_{pl}(R,R))$.
The ratio of the partial cross sections
$ \sigma^{\rmA @\rmC_N}_{ll_0}(\om;\bfa)/\sigma^{\rmA}_{ll_0}(\om)
\equiv \eta_{l}(\om;\bfa)$,
calculated for $\bfa=0$,  reads
\begin{eqnarray}
\eta_{l}(\om; 0)
=
\Bigl|1 + \alpha\, g_{pl}(R,R) \Bigr|^{-2}
\approx
1 - 2\alpha\, {\rm Re}\, g_{pl}(R,R)\, .
\label{CS_central1.9}
\end{eqnarray}
The photon energy $\om$ is related to $p$ 
 via $p=\sqrt{2(\om+I_0)}$.
The first relation in (\ref{CS_central1.9}) is similar to the
formula derived in \cite{AmusiaBaltenkov1998,Baltenkov1999}.
The approximate relation is valid when the cage potential
is treated perturbatively.

For momenta $p$ large enough that $pR\gg 1$, 
one can use the asymptotic formula 
$g_{pl}(R,R) \propto \exp(-\i pR)\sin(pR -\pi l/2 +\delta_l)$.
Then, eq. (\ref{CS_central1.9}) explicitly reveals the oscillatory 
behavior of $\eta_{l}(\om;0)$ as a function of $p$ (or $\om$)
due to the interference of the  
direct and scattered waves.

%
The exact solution of eq. (\ref{wf.3}) with the potential 
(\ref{CagePotential.1}) taken for the {\em off-the-center position}
has not been found.
However, as mentioned, the cage potential can be treated perturbatively.
The first-order solution of (\ref{wf.3}) reads
$\Psi^{(-)}_{\bfp}(\bfr; \bfa) 
\approx \psi^{(-)}_{\bfp}(\bfr) + \Delta \psi^{(-)}_{\bfp}(\bfr; \bfa)$, 
where
\begin{eqnarray}
\Delta \psi^{(-)}_{\bfp}(\bfr,\bfa)
=\!
A\!\!
\int\!\! \d \bfr^{\prime}
G_{\E}^{(-)}(\bfr, \bfr^{\prime}) 
\delta\bigl(|\bfr^{\prime} - \bfa|  -  R\bigr)
\psi^{(-)}_{\bfp}(\bfr^{\prime})
\label{non-central_v2.2}
\end{eqnarray}
is the scattered wave.
The total amplitude of PI is calculated as 
$\calM^{\rmA @\rmC_N}(\bfa)
\!=\! \int\! \d \bfr \Psi^{(-)*}_{\bfp}(\bfr; \bfa)\, \bfe\cdot\widehat{\bfd}
\,\psi_0(\bfr) = \calM^{\rmA} + \Delta\calM(\bfa)$,
where $\calM^{\rmA}$ is the amplitude for the free atom and 
$\Delta\calM(\bfa)$ is due to (\ref{non-central_v2.2}).
Other notations include the unit vector of the photon polarization $\bfe$, 
the operator of the dipole moment $\widehat{\bfd}$, the wavefunction of 
the subshell $\psi_0(\bfr)$.
Squaring the modulus of the amplitude and retaining the terms linear in
the potential, one writes the cross sections as
$\sigma^{\rmA @\rmC_N}(\om;\bfa)\approx \sigma^{\rmA}(\om) + 
\Delta\sigma(\om;\bfa)$, where the term
$\Delta\sigma(\om;\bfa) \propto \Re\bigl(\calM^{\rmA}\,\Delta\calM^{*}(\bfa)
\bigr)$ contains the dependence on $\bfa$ and, thus, must be averaged.
 
First we consider the angular averaging 
$\overline{\sigma^{\rmA @\rmC_N}}(\om;a) \equiv
\int\sigma^{\rmA @\rmC_N}(\om;\bfa)\d \Om_{\bfa}/4\pi$.
The dependence on the direction of $\bfa$ enters 
$\Delta\sigma(\om;\bfa)$ via the
$\delta$-function from (\ref{non-central_v2.2}).
For its averaging one can use the expansion (e.g.,
\cite{VMX}):
$\delta(|\bfr - \bfa|  -  R)
=
(2\pi R/ ar)\sum_{lm} 
P_l(\xi) Y_{lm}(\Om_\bfr) Y_{lm}^{*}(\Om_\bfa)$, 
where $P_l(\xi)$ are the Legendre polynomials, and
$\xi = (r^2 + a^2  -R^2)/ 2ar$ subject to
$|\xi| \leq 1$.
Hence $\overline{\delta(|\bfr - \bfa|  -  R)}=R/ 2ar$ if $|\xi| \leq 1$
and $=0$ if otherwise.
The rest of the algebra is straightforward. 
To evaluate the radial integral in $\Delta\calM(\bfa)$ one 
uses the approximation discussed in connection with eq. 
(\ref{CS_central1.9}).
Finally,  noticing that for each partial transition 
$l_0\to l=l_0\pm 1$ the averaged term $\overline{\Delta\sigma}(\om;a)$
is proportional to $\sigma^{\rmA}_{ll_0}(\om)$, 
one derives the following expressions for 
the ratio 
$\overline{\eta}_{l}(\om;a)=
\overline{\sigma_{ll_0}^{\rmA @\rmC_N}}(\om;a)/\sigma_{ll_0}^{\rmA}(\om)$:
\begin{eqnarray}
\overline{\eta}_{l}(\om;a)
&=&
1 - \alpha
{R \over a}
\int\limits_{R-a}^{R+a}
\d r
{\Re\, g_{pl}(r,r) \over r}
\label{Angular_only1.4}\\
&=&
1 
+ 
 (-1)^{l}
\alpha
\Bigl[
S_p(a)\cos 2 \delta_l
+
C_p(a)\sin 2 \delta_l
\Bigr]
\nonumber
\end{eqnarray}
where  
$S_p(a)=(R/2pa)\int_{R-a}^{R+a}\d r\,r^{-1} \sin (2pr)$
and 
$C_p(a)=(R/2pa)\int_{R-a}^{R+a}\d r\,r^{-1} \cos (2pr)$.

The first relation in (\ref{Angular_only1.4}) {\em qualitatively}  
explains the impact of the angular averaging. 
For $a=0$ the formula reproduces the right-hand side of eq. 
(\ref{CS_central1.9}), derived for the centrally positioned atom 
treated as a point-like source.
As  $a$ increases the averaging leads to a non-zero effective size 
of the source, $D=2a$. 
When $D$ becomes larger than the emitted half-wavelength the
interference is lost.
Hence, for  $p > p_{\min}=\pi/2a$ 
the confinement resonances in the cross section profile disappear.

The right-hand side of (\ref{Angular_only1.4}) allows one to  analyze 
{\em quantitatively} the modification of the interference pattern 
without calculating the parameters of photoelectron wavefunction
(in particular, the phaseshifts)
but using only $p$, $R$ and $a$.
Indeed, the {\em relative} change in the amplitude of the 
oscillations can be understood by comparing the functions
$S_p(a)$ and $C_p(a)$ to their values at $a=0$.

\begin{figure}[ht]
\parbox{4.25cm}{
\includegraphics[clip,scale=0.18]{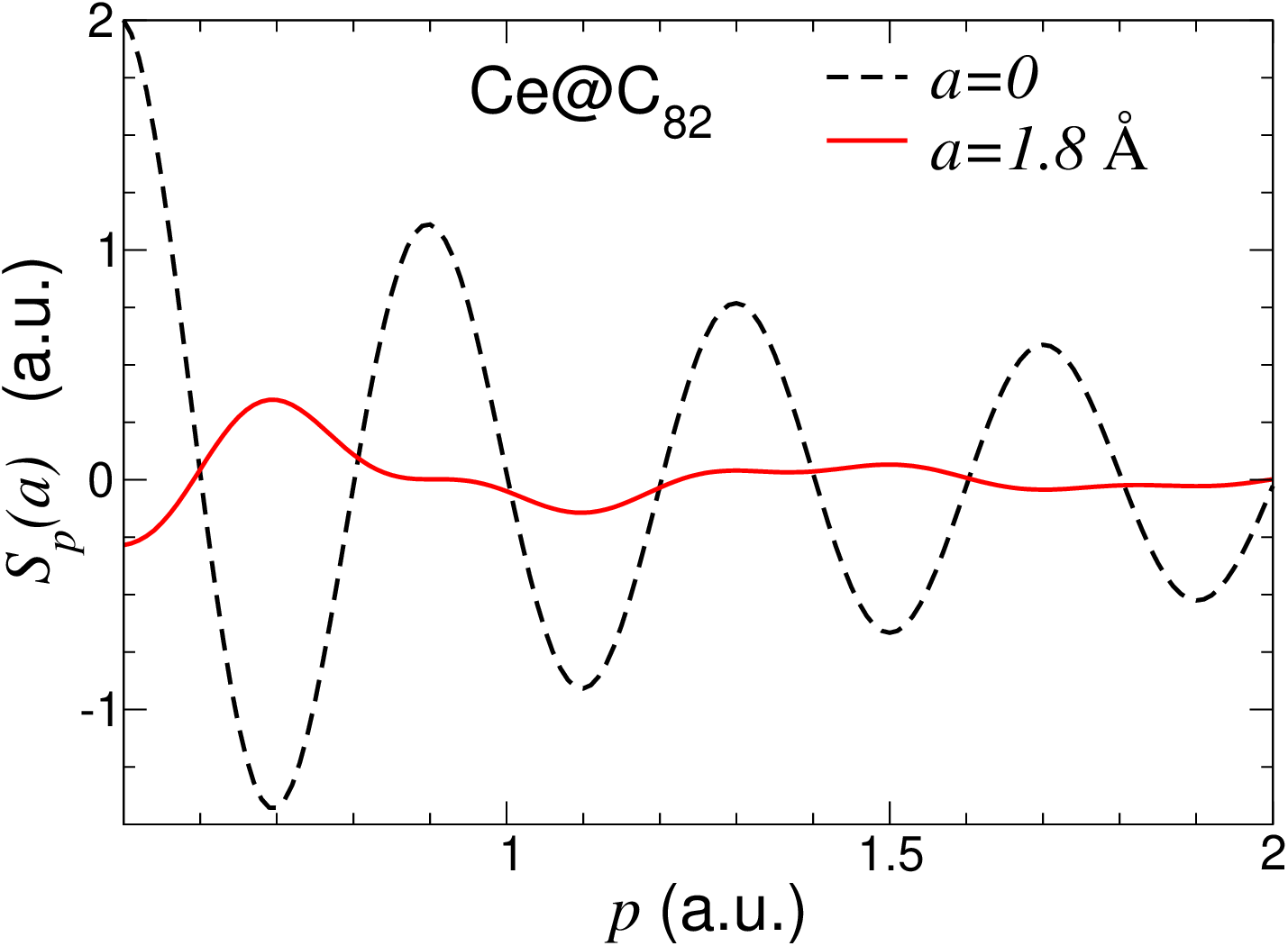}
}
\parbox{4.25cm}{
\includegraphics[clip,scale=0.18]{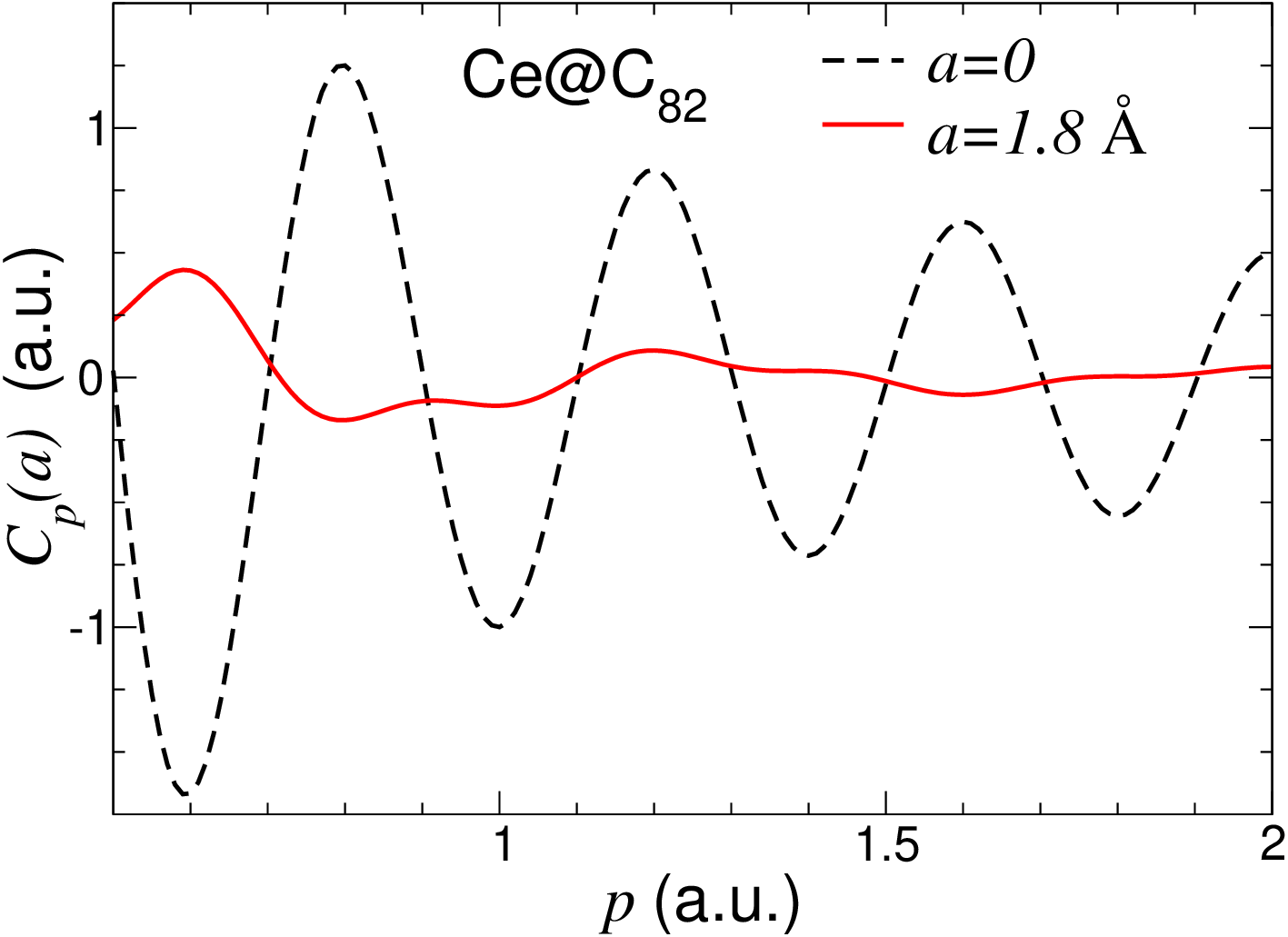}
}
\caption{\!\! 
$S_p(a)$ and  $C_p(a)$ versus $p$ calculated for Ce@C$_{82}$ for 
at-the-center position of Ce and off-the-center with $a\!=\!1.8$ \AA.
}
\label{Ce@C82.fig}
\end{figure}

Let us apply eq. (\ref{Angular_only1.4}) to estimate the effect 
for Ce@C$_{82}$.
In \cite{Mitsuke_EtAl2005,MuellerEtAl2007}  the PI cross section
was measured in the region of the 4d giant dipole resonance, i.e. 
for $\om = 120\dots 140$ eV, and no oscillations were seen.
The 4d ionization potential in Ce is $114$ eV for 4d$_{3/2}$ shell and $111$ eV
for 4d$_{5/2}$  shell \cite{RadzigSmirnov}.
Hence, the indicated range of $\om$ corresponds
the photoelectron momenta  $p = 0.5\dots 1.5$ a.u.
The mean radius of C$_{82}$ is
$4.15$ \AA\, 
and Ce atom is displaced by $1.8$ \AA \, 
from the center 
\cite{Nagase_EtAl1993_Nuttall_EtAl2000}.
The dependences $S_p(a)$ and $C_p(a)$ on $p$ 
are presented in  Fig.~\ref{Ce@C82.fig}.
It is clearly seen that angular averaging destroys the interference
for $p> p_{\min} \approx 0.46$ a.u, i.e. for the 
photon energies $\om \agt I_0 + 3$ eV.
   
Similar estimates  explain the absence of the confinement
resonances for Pr@C$_{82}$ \cite{KatayanagiEaAl_2008}.
Less univocal conclusion can be drawn for 
Li@C$_{60}$, K@C$_{60}$ and Mg@C$_{60}$ the PI of which
was considered theoretically \cite{DeclevaEtAl1999}.
The authors indicate that the atoms displacements are 
1.5, 0.25 and 1 \AA\, for Li, K, and Mg, respectively,
but set $a=0$ when calculating the cross section.
As a result, for each metallofullerene they predicted several 
confinement 
resonances for the photoelectron energies below 2 a.u. 
However, calculating $\E_{\min}=p_{\min}^2/2$ 
and obtaining the values 0.15, 5.5 and 0.35 a.u., we state that
the resonances in  Li@C$_{60}$ and Mg@C$_{60}$ will hardly 
survive the averaging procedure, whereas the resonances predicted
for K@C$_{60}$ will be preserved due to small value of $a$.

To develop the theory further one accounts for thermal 
vibrations of the atom in the vicinity of equilibrium position.
This can be important, e.g., for endohedral noble-gas atoms.
%
Let us construct a model potential for 
the A-C$_{N}$ interaction 
which, despite being quite crude,
allows one to analyze the influence of the vibrations
on the interference pattern.
Ignoring the atom-fullerene hybridization 
one  builds the potential as a sum of pairwise A-C 
Lennard-Jones potentials over all carbon atoms \cite{Albert_EtAl2007}.
Assuming a homogeneous distribution
of carbon atoms over the sphere of radius $R$ and substituting the 
sum with the surface integral,
one derives the A-C$_{N}$ potential 
$U(a)=U_{\rm L}(a)+U_{\ex}(a)$ 
as a function of $a$ \cite{LoKorolSolovyov2009}.
The term $U_{\rm L}(a)$ is the attractive (the London-type) 
potential, 
and $U_{\ex}(a)$ is the repulsive exchange term.
\begin{figure}[ht]
\parbox{4.3cm}{
\includegraphics[clip,scale=0.18]{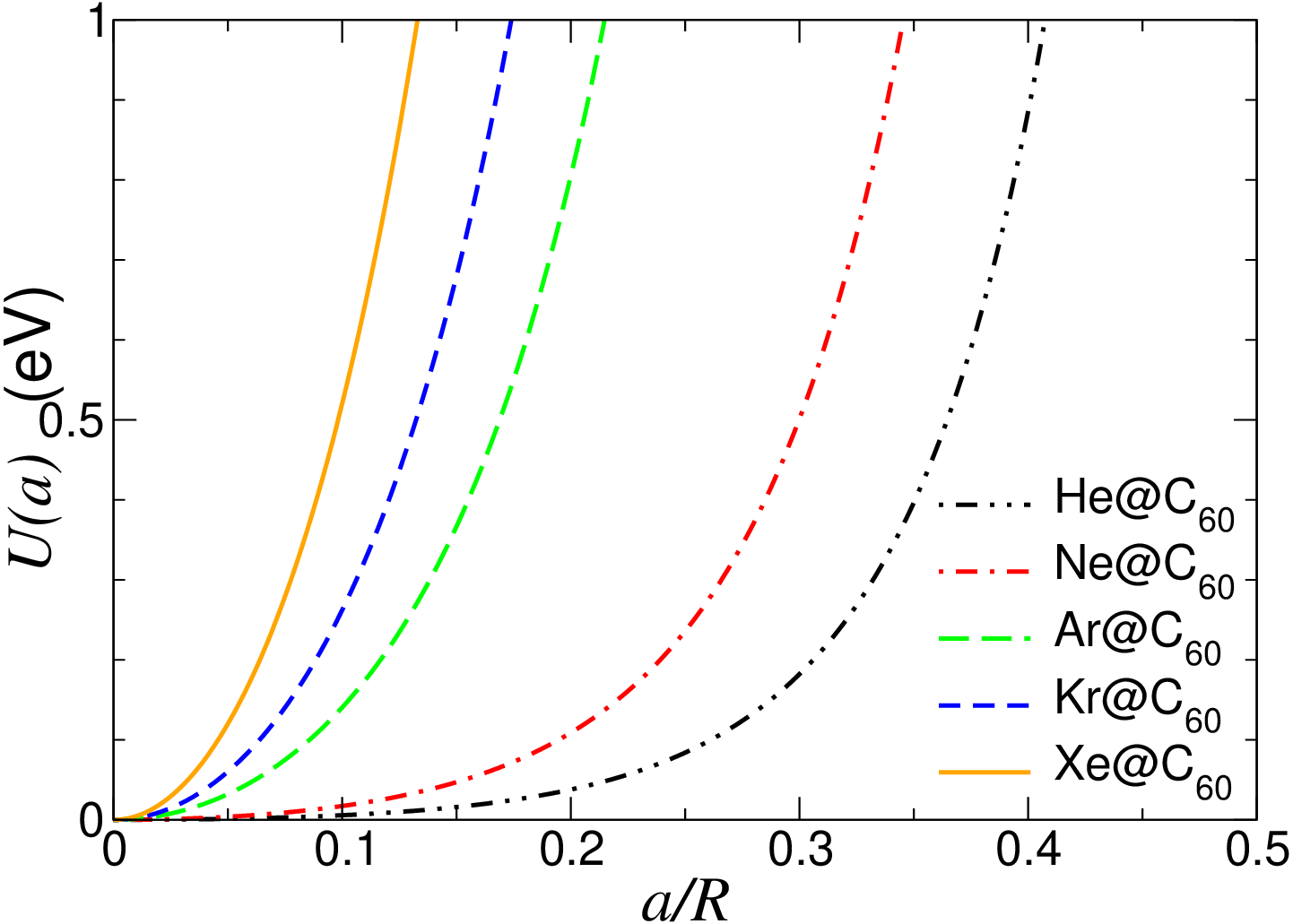}
}%
\parbox{4.3cm}{
\includegraphics[clip,scale=0.18]{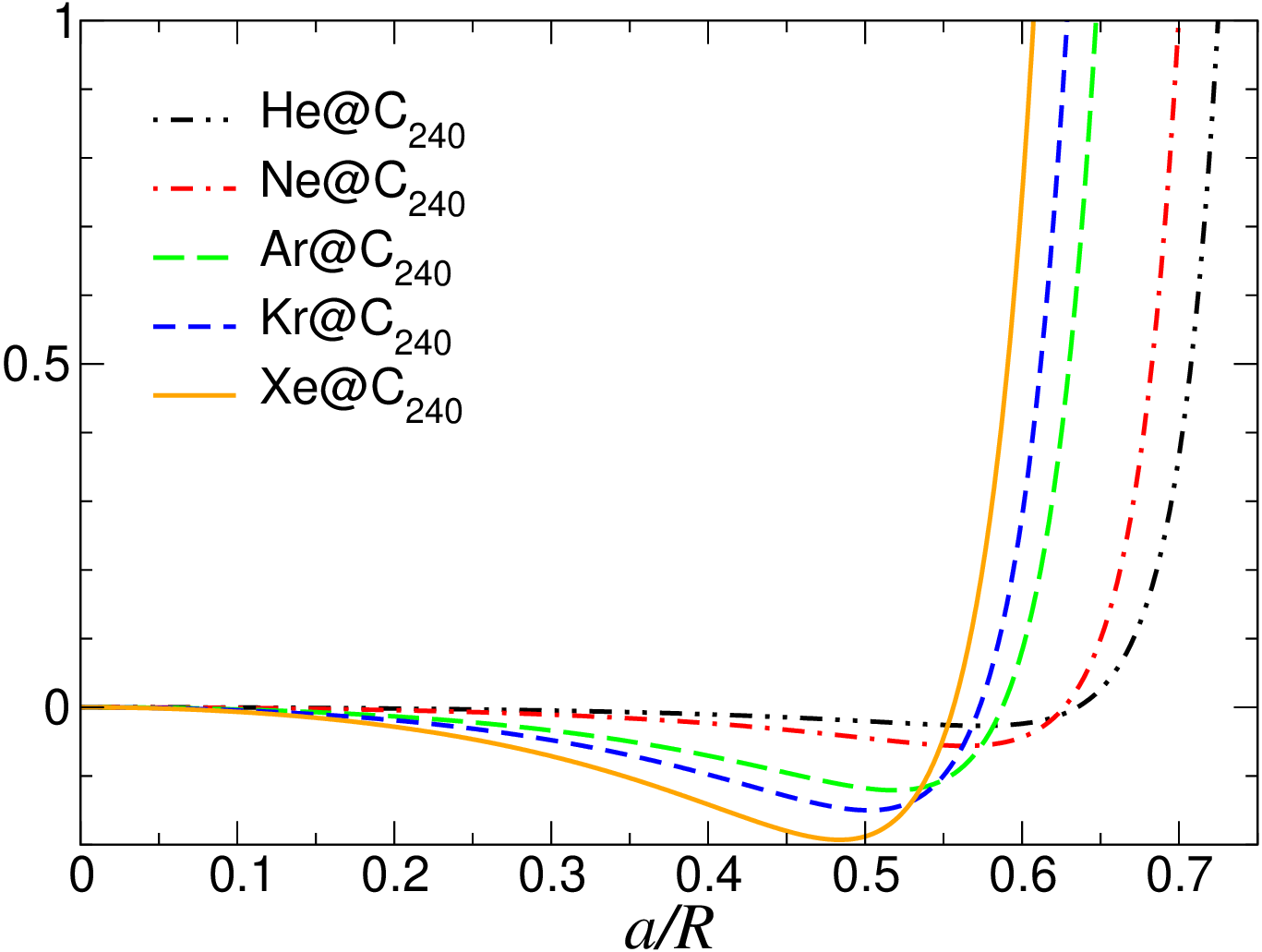}
}
\caption{
Potentials $U(a)$ for various A@C$_{N}$ as indicated.
}
\label{A@CN_pot.fig}
\end{figure}

The functions $U(a)$ for various systems are presented in 
Fig.~\ref{A@CN_pot.fig}.
The radius  $3.5$ \AA\ of the C$_{60}$ cage is less (for Ar, Kr, Xe) 
or just above (He, Ne) the equilibrium distance of the Lennard-Jones 
potential \cite{Albert_EtAl2007}, 
so that the atom is confined to the center \cite{Remark1}. 
For C$_{240}$ the radius $7.1$ \AA\ exceeds the equilibrium distance. 
Hence, the atom is either weakly bound well off the center (Ar, Kr, Xe) 
or moves freely in most part of the fullerene  (He, Ne).

The probability to find an atom at the distance $a$
is given by $\d W_T(a) = C\exp\left(-U(a)/kT\right) a^2\d a$,
where $T$ is the temperature, $k$ -- the Boltzmann constant,
and $C = \left[\int_0^{\amax} \d W_T(a)\right]^{-1}$ with 
$\amax\approx R-R_{\rm A}$ ($R_{\rm A}$ is the radius of the atom).
To average the ratio $\overline{\eta}_{l}(\om;a)$ over $a$
one multiplies eq. (\ref{Angular_only1.4}) by $\d W_T(a)$ and 
integrates over $a$:
\[\langle \overline{\eta}_{l}(\om)\rangle 
=
1 
+ 
\alpha
\left[
\langle S_p\rangle\cos 2 \delta_l 
+
\langle C_p\rangle\sin 2 \delta_l 
\right]
\]
with the general notation 
$\langle g \rangle =\int_0^{\amax} g(a)\,\d W_T(a)$.
%
\begin{figure}[ht]
\parbox{8.4cm}{
\includegraphics[clip,scale=0.3]{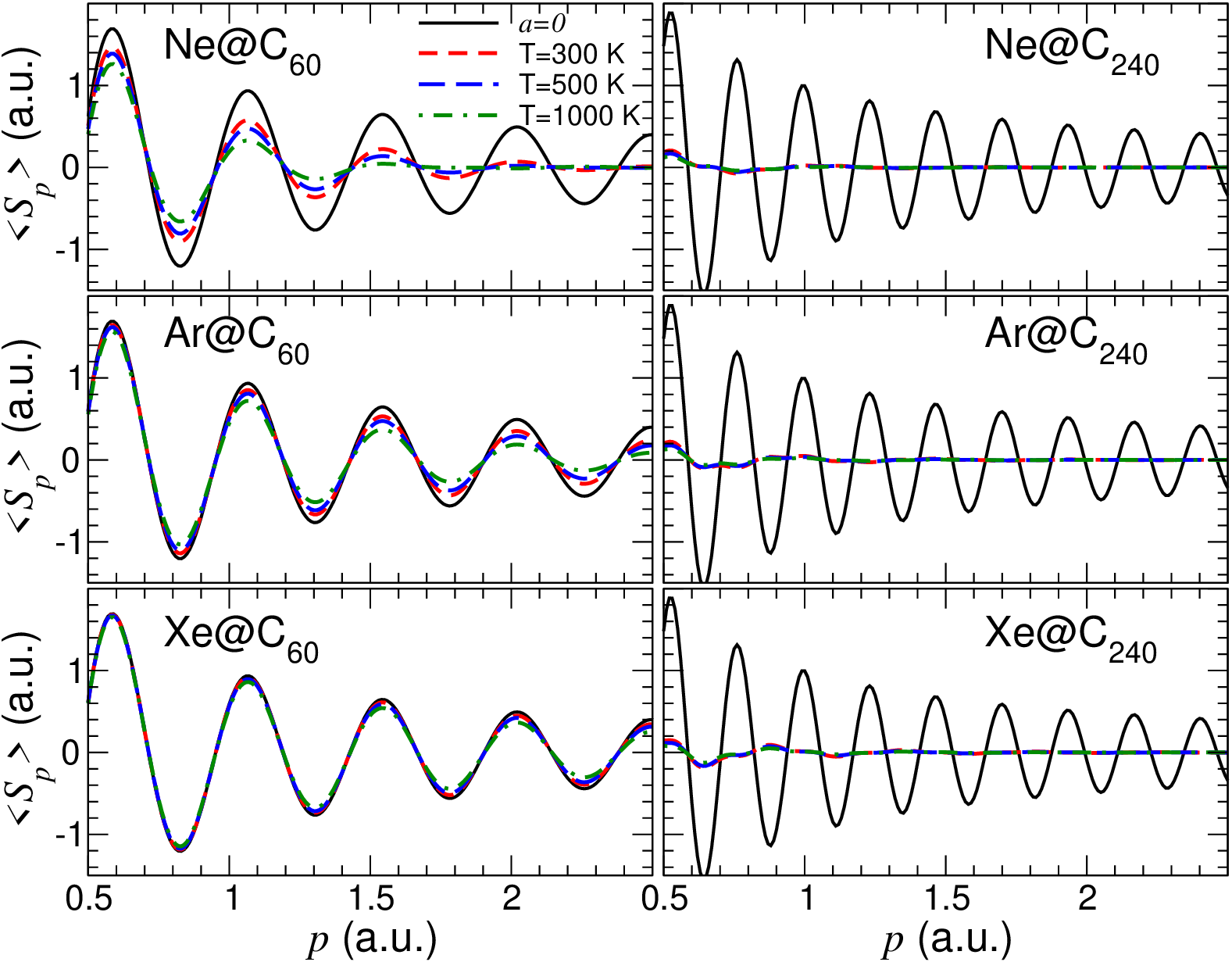}
}
\caption{
$\langle S_p\rangle$ versus $p$ calculated for various 
$T$  and A@C$_{N}$.
}
\label{NG@CN_Sp.fig}
\end{figure}

Dependences $\langle S_p\rangle$ on $p$ are presented in 
Fig.~\ref{NG@CN_Sp.fig} \cite{Remark2}.
It is seen that in all A@C$_{240}$ systems the oscillations are suppressed
because of strong off-the-center position of the atom.
The situation is not so universal for a smaller C$_{60}$ fullerene 
in which the equilibrium  is at the center.
Indeed, for Ne the amplitude of oscillations decreases noticeably for 
all $T$ starting with $p\approx 1$ a.u., whereas for Xe the 
oscillations stay nearly unchanged up to $p\gg 1$ a.u.
To explain the difference one estimates the effective size of the 
source as the doubled mean amplitude of thermal vibrations and 
calculates the momenta $p_{\min}=\pi/D$ beyond which the interference 
disappears.
For $T=500$ K the calculated values of $p_{\min}$ are 1.8, 3.9, 7.1 a.u.
for Ne@C$_{60}$, Ar@C$_{60}$ and Xe@C$_{60}$, respectively.

In summary, we have demonstrated that the interference effects in 
photoionization of A@C$_{N}$ are very sensitive to the mean 
displacement $\langle a \rangle$ of the atom from 
the cage center.
The range of photon energies, where the confinement resonances 
can be observed, one deduces by matching $2\langle a \rangle$
the photoelectron half-wavelength.
In metallofullerenes, due to a non-central position of the atom, 
the angular averaging can destroy the oscillatory structure 
predicted for the atom at-the-center.
For large noble-gas endohedral fullerenes (C$_{240}$ and larger) 
the resonances do not survive the averaging 
procedure with the Boltzmann distribution function.
For a smaller C$_{60}$ fullerene
a more rigorous treatment of the atom-fullerene interaction 
is needed to draw
the final conclusion on the existence of the resonances.
The role of non-centrality must be studies for other 
related phenomena 
(Coulomb confinement resonances, 
non-dipole effects, the photoionization of onion-like structures 
etc \cite{Dolmatov2009}).

The work was supported by the European Commission within the 
NoE project EXCELL (Project No 515703).



\begin{thebibliography}{99}

\bibitem{ConneradeDolmatovManson1999} 
         J.-P.~Connerade, V.K.~Dolmatov, and S.T.~Manson,
         J. Phys. B  {\bf 33}, 2279 (2000).

\bibitem{PuskaNieminen1993} 
         M.J.~Puska and R.M.~Nieminen,  
         Phys. Rev. A {\bf 47}, 1181 (1993); 
         {\bf 49} 629 (1994) (Errata).

\bibitem{WendinWaestberg1993} 
         G.~Wendin and B.~W\"{a}stberg,  
         Phys. Rev. B {\bf 48}, 14764 (1993). 
\bibitem{LuberekWendin1996} 
         J.~Luberek and G.~Wendin,
         Chem. Phys. Lett. {\bf 248}, 147 (1996). 

\bibitem{Baltenkov1999} 
         A.S.~Baltenkov, 
         J.~Phys.~B {\bf 32}, 2745 (1999). 

\bibitem{DeclevaEtAl1999} 
         P.~Decleva  {\it et al.}, 
         J.~Phys.~B {\bf 32}, 4523 (1999). 

\bibitem{AmusiaBaltenkovBecker2000} 
         M.Ya.~Amusia, A.S.~Baltenkov, and U.~Becker, 
         Phys. Rev. A {\bf 62},  012701  (2000).

\bibitem{BaltenkovDolmatovManson2002} 
         A.S.~Baltenkov, V.K.~Dolmatov, and S.T.~Manson,
         Phys. Rev. A {\bf 66},  023201 (2002).

\bibitem{Aichinger2003} 
         M.~Aichinger {\it et al.}, 
         J. Mod. Opt. {\bf 50}, 2691 (2003). 

\bibitem{BaltenkovDolmatovEtAl2003} 
         A.S.~Baltenkov {\it et al},
         Phys. Rev. A {\bf 68},  043202 (2003).

\bibitem{Angular} 
         M.Ya.~Amusia {\it et al}, 
         Phys. Rev. A {\bf 70} 023201 (2004).

\bibitem{AmusiaBaltenkovEtAl2005} 
         M.Ya.~Amusia  {\it et al.}, 
         J. Phys. B {\bf 38},   L169 (2005). 

\bibitem{DolmatovManson2006} 
         V.K.~Dolmatov and S.T.~Manson,
         Phys. Rev. A {\bf 73} 013201 (2006). 

\bibitem{BaltenkovMansonMsezane2007} 
         A.S.~Baltenkov, S.T.~Manson, and A.Z.~Msezane,
         Phys. Rev. A {\bf 76}, 042707 (2007).

\bibitem{DolmatovBrewerManson2008} 
         V.K.~Dolmatov, P.~Brewer, and S.T.~Manson,
         Phys. Rev. A {\bf 78} 013415 (2008).

\bibitem{ChenMsezane2009}
         Z.~Chen  and A.Z.~Msezane,
         J. Phys. B {\bf 42}, 165206 (2009).

\bibitem{Dolmatov2009} 
         V.K.~Dolmatov,
         in {\it Theory of Confined Quantum Sytems: Part 2}, 
         edited by J.R.~Sabin and E.~Br\"{a}ndas, 
         Adv. Quant. Chem. (Academic Press, New York, 2009),
         Vol. 58, p. 13.

\bibitem{Mitsuke_EtAl2005}
         K.~Mitsuke {\it et al.}, 
         J. Chem. Phys. {\bf 122}, 064304 (2005).

\bibitem{KatayanagiEaAl_2008}
         H.~Katayanagi {\it et al.}, 
         J. Quant. Spectrosc. Radiat. Transfer {\bf 109}, 1590 (2008).  

\bibitem{MuellerEtAl2007}
          A.~M\"{u}ller {\it et al.},
          J. Phys.: Conf. Ser. {\bf 88} 012038 (2007).

\bibitem{Shinohara2000} 
         H. Shinohara. 
         Rep. Prog. Phys. {\bf 63}, 843 (2000).

\bibitem{Albert_EtAl2007}
         V.V.~Albert {\it et al.}, 
         Int. J. Quant. Chem. {\bf 107}, 3061 (2007). 
\bibitem{Pyykko_EtAl2007}
        P.~Pyykk\"{o} {\it et al.},
        Phys.Chem.Chem.Phys. {\bf 9}, 2954 (2007).

\bibitem{AtomicUnits} 
         The atomic system of units is used in the paper.

\bibitem{AmusiaBaltenkov1998} 
         M.Ya.~Amusia, A.S.~Baltenkov, and B.G.~Krakov,
         Phys. Lett. {\bf A 243}, 99 (1998).

\bibitem{Messiah} 
        A.~Messiah, 
        {\it Quantum mechanics} (North-Holland,  Amsterdam, 1999).

\bibitem{VMX} 
        D.A.~Varshalovich, A.N.~Moskalev,  and V.K.~Khersonskii,
        {\it Quantum Theory of Angular Momentum}
         (World Scientific, Singapore, 1988).

\bibitem{RadzigSmirnov}
         A.A.~Radzig and B.M.~Smirnov,
         {\it Reference Data on Atoms, Molecules and Ions}
         (Springer, Berlin, 1985). 

\bibitem{Nagase_EtAl1993_Nuttall_EtAl2000}
         S.~Nagase {\it et al.},
         Chem. Phys. Lett. {\bf 201}, 475 (1993); 
         C.J.~Nuttall {\it et al.},
         Mol.Cryst. Liq.Cryst. {\bf 340}, 635 (2000).

\bibitem{LoKorolSolovyov2009} 
         S.~Lo, A.V.~Korol, and A.V.~Solov'yov,
         Phys. Rev. A {\bf 79} 063201 (2009).

\bibitem{Remark1} 
         The model ignores the plasmon 
         excitations in C$_N$, 
         which enhance $U_{\rm L}(a)$ 
         \cite{Pyykko_EtAl2007} and make the curve $U(a)$
         less steep.

\bibitem{Remark2} 
         General trends of $\langle C_p\rangle$ are similar.
         To save space we do not include the corresponding graphs.


\end{thebibliography}
\end{document}